\documentclass[preprint]{aastex}
\usepackage{psfig}
\received{}
\revised{}
\accepted{}
\ccc{}
\cpright{}{}

\slugcomment{Submitted to ApJ}
\shorttitle{X-ray Trifid}
\shortauthors{}

\newcommand{\chandra}{\textit{Chandra}}

\begin{document}

\title{\chandra\ Observation of the Trifid Nebula:
X-ray emission from the O star complex
and actively forming Pre-main sequence stars }

\author{Jeonghee Rho and Solange V. Ram\'{\i}rez }
\affil{SIRTF Science Center, Mail Stop 220-06,
California Institute of Technology.}

\author{Michael F. Corcoran\altaffilmark{1} and Kenji Hamaguchi}
\affil{Code 662, NASA/Goddard Space Flight Center, Greenbelt, MD 20771}
\altaffiltext{1}{Universities Space Research Association, 
7501 Forbes Blvd, Ste 206, Seabrook, MD 20706}
 
\author{Bertrand Lefloch}
\affil{Laboratoire d'Astrophysique, Observatoire de Grenoble, BP 53,
F-38041, Grenoble CEDEX9, France}

\begin{abstract}
The Trifid Nebula, a young star-forming HII region, was
observed for 16 hours by the ACIS-I detector on board of the  \chandra\
X-ray Observatory. We detected 304 X-ray sources, thirty percent of
which are hard sources and seventy percent of 
of which have  near-infrared counterparts. \chandra\  resolved the HD164492 multiple system
into a number of  discrete  X-ray sources. X-ray emission is detected
from components HD164492A (an O7.5III star which ionizes the  nebula),
B and C (a B6V star), and possibly D (a Be star). Component C is  blended
with an unidentified source to the NW. HD164492A
has a soft spectrum ($kT \approx 0.5$ keV) while the component C
blend shows much harder emission ($kT\approx$ 6 keV). This blend and
other hard sources are responsible for the hard  emission and Fe K line
seen by \emph{ASCA}, which was previously  attributed entirely to
HD 164492A. The soft spectrum of the O star is similar to emission seen
from  other single O stars and is probably produced by shocks  within
its massive stellar wind.  Lack of hard emission suggests that neither
a magnetically  confined wind shock nor colliding wind emission is
important in HD164492A.  
 A dozen  stars  are found to have flares in
the field and most of them are pre-main sequence stars (PMS).  Six
sources with  flares have both optical and 2MASS counterparts.  These
counterparts are not embedded and thus it is  likely that these sources
are in a later stage of PMS evolution, possibly Class II or III. Two
flare sources did not have any near-IR, optical, or radio counterparts.
We suggest these X-ray flare stars are in an early pre-main sequence
stage (Class I or earlier). We also detected X-ray sources apparently
associated with two massive star forming cores, TC1 and TC4. The
spectra of these sources show high extinction and X-ray luminosities of
$2 - 5\times 10^{31}$ erg s$^{-1}$. If these sources are Class 0
objects, it is unclear if their X-ray emission  is due to solar-type
magnetic activities as in Class I objects, or some other mechanism.
\end{abstract}


\section{Introduction}

Despite of the fact that X-ray observations of star-forming regions
have been a powerful tool for discovering young stellar objects (YSO)
and T Tauri stars (TTS) since  {\it Einstein}  (references in Feigelson
\& Montmerle 1999), long wavelength (IR and submillimeter) emission is
used to define evolutionary classes for young stars. Classes I-III are
solely based on the excess seen in the IR spectral  energy distribution
(SED) with respect to stellar blackbody photospheric emission, as
measured by the spectral index  $\alpha_{IR}$ = d log ($\lambda$
F$_\lambda$) / d log $\lambda$ between $\lambda$ = 2.2 and 10-25$\mu$m:
Class I , II and III correspond to $\alpha_{IR}$$>$ 0, -2
$<$$\alpha_{IR}$ $<$ 0, and   $\alpha_{IR}$ $<$ -2 (Lada 1987). Class 0
stars represent the earliest phase of star formation and are
identified by the ratio of submillimeter to bolometric luminosity
(Andr\'e et al. 1993) and are seen as condensations in submillimeter
far-infrared dust continuum maps; these condensations often show
collimated CO outflows  or internal heating sources. X-ray bright T
Tauri stars usually belong to either Class II or III.

High resolution images obtained by the  \chandra\ X-ray Observatory
open a new era in the study of star formation because the 1$''$
resolution of  \chandra\ is necessary to resolve individual sources
in nearby star forming regions, and because confusion  due to foreground and
background stars is much less important at X-ray energies than in the
optical and near-infrared regime. Recent {\it Chandra} observations of Orion
\citep{gar00,fei02}  detected hundreds of X-ray sources, e.g. pre-main
sequence stars (PMS) with masses in the range 0.05 M$_{\odot}- 50$
M$_{\odot}$ and a combined infrared and X-ray study suggests that the
X-ray luminosity of PMS depends on stellar mass, rotational history,
and magnetic field \citep{gar00}. A high percentage of Class I  PMS
were also found to be  X-ray emitters; in $\rho$ Oph,  70\% of
identified Class I stars  are X-ray bright.
Strong X-ray flares from PMS in  $\rho$ Oph \citep{ima01}, Monoceros
R2 \citep{koh02}, and Orion \citep{fei02} were detected from Class I, II
and III objects, possibly because of magnetic activity.  Moreover,
X-ray emission from Class 0 candidates  was detected in OMC-3
\citep{tsu01}.  

The Trifid Nebula (M20) is one of  the best-known astrophysical
objects.   It is a classical nebula ionized by an O7.5 star,  HD164492A,
and the ionized nebula  glows  in red light.  The nebula is trisected by
obscuring  dust lanes giving the Trifid its name. A blue reflection
nebula appears to the north of the red nebula.  At an age of $\sim
3\times 10^{5}$ years, the Trifid is a young
\ion{H}{2} region. Recent studies using the Infrared Space Observatory
(ISO) and the Hubble Space Telescope (HST) (Cernicharo et al.\ 1998;
Lefloch \& Cernicharo 2000; Hester et al.\ 1999) show the Trifid to be
a dynamic, ``pre-Orion" star forming region containing young stars
undergoing episodes of violent mass ejection, with protostars like
HH399 \citep{lef02} losing mass  and energy to the nebula in optically
bright jets. Four massive (17--60 M$_\odot$) protostellar cores were
discovered in the Trifid from millimeter-wave observations.
These cores are associated with molecular gas condensations at the edges of
clouds (Lefloch \& Cernicharo 2000).  T Tauri stars  and  young
stellar object candidates were also identified using near-infrared
color-color diagrams from 2MASS data  (Rho et al. 2001). Unlike
better-studied nearby star-forming regions such as Orion and $\rho$
Oph,  star forming activities in the Trifid have only recently been
recognized \citep{cer98},  and  as a result of this (and also due
to contamination by foreground and background  stars in the optical
and IR), the population of PMS in the Trifid has not been fully
investigated.
The distance to the Trifid Nebula is between 1.68 and 2.84 kpc (Lynds et al.\ 1985;
Kohoutek et al.\ 1999 and references therein) and 
a distance of 1.68 kpc is adopted in this paper.

In this paper, we report  results of \chandra\  observations of the
Trifid Nebula. \chandra\ resolved the HD 164492 multiple system
into a number of discrete X-ray sources and we present 
their X-ray properties which include components A (an O star),
C (a B6V star) and B (A2Ia).
Our \chandra\ observations revealed 304 sources and 
we found  that 30\%  of the sources have hard emission similar to that from
PMS.   Among these candidate PMS, we  report  properties of about a
dozen flare  sources, which include unusual variability from the O
star and  unusual emission from an A-type supergiant.  We also discuss
the X-ray properties of the HD 164492 complex, the properties  of X-ray
sources which are  apparently associated with two protostellar cores, and
also the properties of some apparently strongly variable  objects.

\section{Observations}

The Trifid Nebula was observed with the Advanced CCD Imaging
Spectrometer (ACIS) detector on board the {\it Chandra X-ray
Observatory} \citep{wei02} on 2002 June 13.  The results presented here
arise from the imaging array (ACIS-I), which consists of four
1024$\times$1024 front-side illuminated CCDs. The array was centered at
R.A.\ $18^{\rm h} 02^{\rm m} 28^{\rm s}$ and Dec.\
$-22^\circ$56$^{\prime} 50^{\prime \prime}$ (J2000) and covered an area
in  the sky of about $17'\times17'$. The total exposure time of the
ACIS observations was 58 ksec. This observation is sensitive to X-ray
luminosities of 5$\times$ 10$^{29}$ erg s$^{-1}$, assuming an
appropriate PMS X-ray spectrum (temperature of 1 keV and an absorption
1.6$\times$ 10$^{21}$ cm$^{-1}$) at the distance (1.68 kpc) of the
Trifid.

We started data analysis with the Level 1 processed event list provided
by the pipeline processing at the {\it Chandra} X-ray Center.  The
energy and grade of each data event were corrected for charge transfer
inefficiency (CTI), applying the algorithm described by \citet{tow00}.
The event file was filtered to include event grades of 0, 2, 3, 4, and
6, and filtered by time intervals to exclude background flaring
intervals or other bad times. The filtering process was done using the
\chandra\ Interactive Analysis of Observations (CIAO)
package\footnote{http://cxc.harvard.edu/ciao/index.html}  provided by
the {\it Chandra} X-ray Center.

Figure ~\ref{chandrazoom} shows the \chandra\ ACIS-I three color image
of the Trifid. Hundreds of point sources are detected with little
diffuse emission.  X-ray sources were located using the {\it wavdetect}
tool within the CIAO package. This tool performs a Mexican hat wavelet
decomposition and reconstruction of the image after accounting for the
spatially varying point spread function as described by \citet{fre02}. 
We used wavelet scales ranging from 1 to 16 pixels in steps of
$\sqrt{2}$, and a default source threshold probability of
$1\times10^{-6}$. The {\it wavdetect} tool was run using an exposure
map and it produced a  catalog of 353 sources from the entire ACIS-I
FOV. Then we identified false sources produced by cosmic rays or  cases
in which the source counts are below the background counts.   This
observation  finally resulted in 304 X-ray sources detected from the
total ACIS-I FOV ($17'\times17'$).  The  full source list is given
in Table ~\ref{catalog} in order of R.A. 
Thirty percent of these sources are shown to be
hard (shown in blue in Figure ~\ref{chandrazoom});
 the hard sources
 have a spectral hardness ratio (SHR) $> -0.2$, where SHR is the ratio
of the net counts in the hard $2.0-8.0$ keV band to those in the soft
$0.5-2.0$ keV band.
Diffuse emission was not obvious in the \chandra\ images 
of the Trifid Nebula with our 
current data processing. None-detection of diffuse emission  
in M20 
is consistent with a claim
that high mass star-forming
regions without stars earlier than O6 may be unlikely to exhibit diffuse
soft X-rays (Townsley et al.\ 2003; Abbott 1982)

The brightest  source in the field corresponds to the O star HD164492A
(source 102) and has 884 counts. This is equivalent to 0.047 photons per frame,
which  is small enough so as not to be affected by pileup effects (the
pileup fraction  is $<< 0.05$).  In order to verify the hard  emission
and Fe K line detected in the \emph{ASCA} spectra of HD164492A  (Rho et al.
2001), we extracted a spectrum using the same extraction region as was
used in the  \emph{ASCA} analysis, and confirmed that the \chandra\ and
\emph{ASCA} spectra are the same within the errors. However, we found
that HD164492A is actually a  soft source (see section 6  for details)
while the hard emission and Fe K line  (which were attributed to the O
star in the  \emph{ASCA} analysis)  is actually produced by the hard
sources resolved by \chandra\ as shown Figure  \ref{ostarcomplex}. In
addition, the emission previously attributed entirely to HD164492A in
an analysis of \emph{ROSAT} PSPC  data (Rho et al. 2001) is now
resolved into a dozen X-ray sources as shown in Figure
\ref{chandrazoom} and \ref{ostarcomplex}.

To confirm the X-ray astrometry, we used the 
2MASS\footnote{http://www.ipac.caltech.edu/2mass} final release point
source catalog to search for near-IR counterparts to the X-ray sources
in the full source list. We
cross-correlated the positions of the 2MASS sources with the positions
of our X-ray sources. Within $5'$ of the \chandra\ aimpoint, we found
72 2MASS sources that coincide with X-ray source positions. We computed
the total R. A. and DEC offset of these sources and obtained offsets of
$-0.16"$ in R. A., and $-0.09"$ in DEC,  which are less than one third
of a pixel. Therefore, the  astrometry of Trifid ACIS image was
confirmed with the minor systematic errors in the {\it Chandra} aspect
solution.

\section{The Multiple System HD164492}

The HD164492 complex is a multiple stellar system composed of 7 
physically related components, A-G  \citep{koh99}.  Component HD164492A
is the  O7.5 III((f)) star responsible for ionizing the nebula.  This
complex  was observed as an unresolved bright X-ray source in PSPC
images (Rho et. al. 2001). Our \chandra\ images (Figures
\ref{chandrazoom} and \ref{ostarcomplex}) resolve this complex into a
number of discrete X-ray sources. X-ray emission is clearly detected
from components A, B, and C as shown in Table 2 (Fig.
\ref{ostarcomplex}). HD164492A is the brightest X-ray source in this
complex. Component HD164492C  (a B6V star, Gahm et al.  1983) is a bright X-ray
source (source 94) and is barely  resolved from a nearby X-ray source to the NW 
of C (source 91) which has no  obvious optical counterpart (Fig.
\ref{ostarcomplex}).  Source D (a Be star) is a  strong H-$\alpha$
emitter (Herbig 1957) and has a disk (resolved by a radio image), similar
to proplyds in the  Orion nebula (O'Dell 2001). The component D is much
fainter  in X-rays than component C.

\subsection{HD 164492A}

The X-ray light curve of the HD164492A shows small but significant
variability ($\sim 20\%$ in the 10 hour observation, Figure
\ref{lcpms}).  This is unsual since most OB stars are not known X-ray
variables. X-ray variability is not widely known from early type stars
based on previous \emph{Einstein} and \emph{ROSAT}  surveys
\citep{chleb89,ber96}. However,  recent \chandra\  observations show
significant variability from some early type stars \citep{fei02}. For
example, $\theta^2$ OriA, an O9.5 star, shows a 50\% drop in 10 hr with
multiple 10-20\% flares \citep{fei02}. Some level of  time variability
is also observed from other early type stars in Orion \citep{sch01}. 
The spectrum of the O star
shows thermal emission as shown in Fig. \ref{ostarspec}a.
We fit the X-ray spectra of 
HD164492A using two collisional ionization
equilibrium (CIE) thermal models (a Mewe-Kaastra plasma model (Kaastra
1992) and an updated Raymond-Smith model (Raymond \& Smith 1977); MEKAL
and APEC in XSPEC). Both models gave similar results. We assumed
sub-solar metal abundance (0.3 solar); assuming solar abundances
produced less than a 20\% change in the derived spectral parameters.
The best fit of the spectrum of O star yields $N_H = 1(<5) \times
10^{21}$ cm$^{-2}$ and a temperature of $kT_s = 0.5 (<0.6)$ keV, which
implies an X-ray luminosity of 2.8$\times 10^{31}$ erg s$^{-1}$.  The
derived temperature is similar to the  X-ray temperatures of other
single massive stars (Corcoran et al. 1994; Moffat et al. 2002).
Although some of the luminosity may be due to an unresolved low-mass 
companion such as a T-Tauri star which could be responsible for the 
variability, the low temperature obtained from the spectral fit is largely 
consistent with a single massive star.
Assuming a bolometric luminosity of L$_{bol}$
$\sim$0.5--1.6$\times10^{39}$ ergs s$^{-1}$, the ratio of X-ray to
bolometric luminosity is  $-7.76 < \log L_{\rm x}/L_{bol}<-7.25$,
which is smaller than the canonical value $L_{x}/L_{bol} \approx -7$
but it is within the scatter \citep{ber96}.  The lack of high
temperature X-ray emission  in HD164492A  suggests that a  magnetically
confined wind shock has not developed in the O star of HD164492A as has
been suggested in  other variable O stars showing high temperature
X-ray emission (like  $\theta$ Ori A, C, and E \citet{sch01}).
Similarly the lack of hard emission in HD164492A  also suggests that
colliding wind emission (as considered in  \citet{rho01}) is not
important.  X-rays from HD164492A are likely produced by shocks
distributed through the star's massive  stellar wind.  The variability
may be from instabilities within a  radiatively driven wind (Lucy \&
White 1980; Feldmeier et al. 1997), but if so  the rapid timescale of
the changes ($\sim 2$ hours) suggests that the  X-ray emission is
dominated by a small number of strong shocks rather  than a large
distribution of weak shocks.

\subsection{HD164492C}

We also modelled the extracted spectrum from source C, which is 
blended with the unidentified X-ray source to the NW (source 91:
Component C2) which is barely  resolved from source C. The component D,
a Be star, is not identified as a source by wavelet analysis; however,
there is some faint X-ray emission  coinciding with the Be star as
shown in Figure 2.  Component C2 (source 91) shows variability, and the
light curve is shown in Fig. \ref{lcpms}.  The best two-temperature
absorbed thermal fit in Fig. \ref{ostarspec}b has $N_H(1) =
7.8^{+5.2}_{-4.8} \times 10^{21}$ cm$^{-2}$ and a temperature of $kT_1
= 0.6\pm0.4$ keV, and $N_H(2) = 1.6(^{+4.4}_{-1.1})\times 10^{21}$
cm$^{-2}$ and a temperature of $kT_2 = 5.9 (^{+\infty}_{-3.4})$ keV.
The luminosity of soft and hard components are 1.4$\times 10^{32}$  and
6$\times  10^{31}$ erg s$^{-1}$, respectively. The lower absorption 
$N_H(1) =  1.6\times 10^{21}$ cm$^{-2}$ is comparable to the optical
extinction $A_{v} = 1$ mag \citep{koh99}, suggesting that the wind(s)
and the strong ionizing radiation from the O and B stars strip the
surrounding materials.  The 6 keV component is  partially responsible
for the spectra seen by \emph{ASCA} (Rho et al. 2001). The X-ray
emission from B and Be stars is not well characterized at present since
they are  often faint.  The $L_{\rm x}/L_{Bol}$ ratio is below
10$^{-7}$$\sim$  10$^{-5}$ for  stars of spectral type B1.5(III-V) and
later \citep{fei02},  and the late type stars have a luminosity range
of  10$^{29}$-10$^{31}$ erg s$^{-1}$.   Thus the observed X-ray
emission of 2$\times 10^{32 }$erg s$^{-1}$  from the source C blend is
difficult to reconcile with emission from a typical B6V type star or a
late type main-sequence companion star.  We extracted separate spectra
from both components C and C2, but the spectra could not be
distinguished from each other and additionally we lose the hard
component emission due to lack of photon statistics in the hard energy
band, although separate light curves show stronger time variability in
the component C2.  The B6V star is an X-ray source as shown in Figure
~\ref{ostarcomplex},  and the component C2 is a  companion  candidate
which is also strong X-ray emitter. The companions of later B-type
stars with strong X-ray emission, are suggested to be often PMS (Stelzer et
al.\ 2003), which show characteristics of high X-ray luminosities and
hard X-rays.   The presence of an unidentified X-ray source near the B6V
star and the presence of hard X-ray emission in this system suggest 
that C2 may be a PMS.  Other possibilities such as emission from
corona of the Herbig Be star,  or a mechanism related to the proplyds
remain open at this point.

\subsection{HD 164492B}

Component B (source 106) is classified as A2Ia (Gahm et al. 1983) or possibly as 
an A2 III (Lindroos 1985).  A-type stars are not often detected as 
X-ray sources \citep{guill99} since they lack subsurface convection 
zones to power strong magnetic fields, and they lack a strong UV flux 
to power massive stellar winds.  This may indicate that this star has 
a low-mass, X-ray bright companion or the spectral type may be 
incorrect. Its X-ray luminosity is listed in Table 2;  we assumed
a temperature of 0.15 keV from other A stars 
(Simon \& Drake, 1993) and the absorption of 
1.6$\times$10$^{21}$ cm$^{-2}$.

\section{Pre-main Sequence star candidates}

We created lightcurves for each of the detected X-ray sources, using  a
bin length of 2500 sec, which gives a total number of 24 bins.   We
calculated the ratio between max count in the highest bin and minimum
count in the lowest bin and found  $\sim$40 sources have the ratio
greater than 1 after accounting for the errors.  If a source was only
detected during a flare, we used the mean count rate instead of minimum
count rate.

In Table 2 we list sources in which this ratio greater than 3 (sources
8, 23, 97, 166, 170, 194, 211, 237, 246, 256, 283 and 285). 
These sources should represent the most  strongly variable
sources.  To confirm this variability we also  performed a simple
$\chi^2$  test  against an assumed the constant light curve.  The
probability of constancy for each source from this test  is given in
Table 2, for which we  used additional binning as needed for each
source.  Nine sources (sources 23, 91, 97, 166, 194, 237, 256, 283
and 285) in Table 2 are definitely variable (null probability $<$
10\%) based on  the $\chi^2$ test and our ratio test. We show  sample
light curves of the variable sources in Fig. \ref{lcpms}.

We searched for near-infrared and optical counterparts for these
variable sources.  Unfortunately no previous optical identification of
T Tauri stars exists in the Trifid. The only  identified T Tauri stars
or young stellar objects  are 85 PMS candidates identified from 
infrared excesses  using 2MASS second release data \citep{rho01}.  Here
we re-evaluate the identification of PMS candidates --TTS or YSO-- 
from near-infrared color-color diagrams  using the 2MASS final release
data. The final release data  have improved photometry by allowing
multiple-components of point spread function for the  blended sources
with the improved zero point, and a complete source catalog by
thoroughly identifying artifacts. Among variable sources in Table 2, only 
source 283 was previously identified as a YSO \citep{rho01} and now
sources 23, 97, 166, 194 and 237 are identified as TTS or YSO based on the
2MASS data. We also searched for optical counterparts to these X-ray
sources using the Guide star catalog (GSC 2.2) and USNO-B catalog, 
but since these catalogs do not
include any sources near bright optical diffuse emission, we also
directly compared the X-ray sources  in Table 1 to  digital all  sky
images (DPOSS). Among the variable sources listed in Table 2, five
(source 97, 166, 237, 256 and 285) have SHR $> -0.2$ (and they appear blue in
Figure \ref{chandrazoom}).

Sources 8, 23, 194, 211, 246 and 283 have both optical and 2MASS
counterparts.  These sources are not embedded and thus are likely in
later stage of PMS evolution, possibly Class II or III.   Sources
8, 211 and 246
have 2MASS counterparts, but not identified as TTS or YSO from
near-infrared colors.  They are still likely PMS,  because the near-IR
excess is shown to be time-dependent (Carpenter et al.
2000). Sources 97, 166, 237, 256 and 285 have no optical counterparts and
have SHR $>-0.2$ and the light curve of source 166
is shown in Fig. \ref{lcpms}. In particular, source 256 and 285 have neither near-IR
nor optical counterparts but exhibit flares in their X-ray light 
curves 
as shown in Fig. \ref{lcpms}. No radio counterparts are
known for these sources. We suggest these X-ray flare stars are in
an early pre-main sequence stage (Class I or earlier). Source 211 has
sufficient counts for spectral fitting, and the best-fit using a
one-temperature thermal model yields $N_H = 1(<8) \times 10^{21}$
cm$^{-2}$ and a temperature of $kT = 2$ ($>$1) keV. The inferred X-ray
luminosity of 1.1$\times 10^{31}$ erg s$^{-1}$, is comparable to those
from TTS and YSO \citep{fei99}, and brighter than those of typical
low-mass main-sequence stars  ($\sim$10$^{30}$ erg s$^{-1}$). The
sources with flares seem to be  found preferentially concentrated along
the dust lane and at the edge of the HII region, i.e. along the
ionization fronts.  This reinforces the conclusion that most of flaring
sources are PMS. We estimated the X-ray luminosities of sources 
8, 23, 97, 166, 170, 194, 246, 256 and 283 assuming the spectral 
parameters are the same as those of source 211.
For sources 237 and 285, we assumed the spectral parameters are the same
as those of TC1 source, because their hardness ratios are comparable to
that of TC1.

\section{Identification of X-ray Emission from Massive Protostellar Cores}

We compared the X-ray sources with the known massive star forming cores
detected in the 1300$\mu$m dust continuum map \citep{lef01}. There are
five such cores, four of which are known to  have bipolar wings. We
detected X-ray emission from  TC1 (source 117) and possibly TC4 (source 38),  two cores with
bipolar wings and associated Class  0 candidates \citep{cer98,lef00}.
The sources associated with both TC1  and TC4 are hard X-ray sources
(which appear as blue sources in Figure \ref{chandrazoom}). Figure
\ref{tc1tc4} shows the positions of the X-ray sources superposed on the
1300$\mu$m dust continuum map. The X-ray source associated with TC1 has
no 2MASS counterpart. But the X-ray identification of the source
associated with the TC4 core is less  clear, because this source is
located at the edge of the ACIS field of view where the width of PSF is
$\sim12''$.  The TC4 source coincides with the 1300$\mu$m dust
continuum source within $11-12''$, and also has two 2MASS counterparts
within $12''$. The closest 2MASS source is 2MASS180212.5-230549 for
which J band emission is not detected and H(=14.8mag) and K$_s$
(=12.9mag) magnitudes are contaminated by a nearby star.   It is not
clear if this 2MASS source is related to the TC4 and the X-ray source.
TC4 satisfies the Class 0 definition of \citet{and93} in terms of its
dust temperature ($\sim 20$ K), the ratio of its submillimeter to
bolometric luminosity, and the presence of a bipolar outflow. An
outflow from TC4 shows a kinematic  age of 6.8$\times$10$^3$ yr and
suggests it to be an intermediate- or high-mass object \citep{lef02}.
For TC1, the  SED is not available so it's not clear if TC1 satisfies
the Class 0  criteria  or not; however, the object is deeply embedded
in a dust core so it is at least a Class I object, and  possibly younger
than a Class I source.

 TC1 and TC4 have
hardness ratios of $0.24$ and $0.67$, respectively (see Table 1 and 2). The
source associated with TC4 is  one of the hardest sources in Table 1.
These sources are either highly absorbed or have high temperatures. Our
analysis of the X-ray spectra of TC1 and TC4 shows that the best  fit
thermal model parameters are N$_H$=6$\times 10^{22}$ cm$^{-2}$,  $kT =
1.7$ keV and $L_{\rm x} = 4.7\times 10^{31}$ erg s$^{-1}$ for the TC 4
source, and $N_H=3.4\times 10^{22}$ cm$^{-2}$, kT = 1.1 keV, and
$L_{\rm x} = 1.9\times 10^{31}$ erg s$^{-1}$ for the TC1 source.  The
large  column densities we derive show that these sources are highly
embedded.
The only detection of X-ray emission from a  Class 0 object until now
is  a source located in OMC-3 (CSO6) \citep{tsu01}.
In comparison, the Class 0 source in OMC-3 (CSO6) has $N_H =
(1-3)\times 10^{23}$ cm$^{-2}$ and a luminosity of $10^{30}$ erg
s$^{-1}$.
X-ray emission from Class 0 objects which are
in the dynamical infall phase is poorly understood since so far only a few
Class 0 candidates have been identified  and their properties are not
well constrained. It is unclear if the X-ray emission is due to
solar-type magnetic activities as in Class I objects.
The fact that these two counterparts have high hardness ratios
does not support the X-ray emission is from a low-mass companion 
to the protostellar core. If these objects 
are  accreting, then the X-ray  emission  may be related to the
accumulation and release of angular momentum toward the growing
central star by outflow processes.

\section{Summary}
Young HII regions like the Trifid are rich sources of
X-ray emitters.  Our \chandra\ images reveal a few hundred X-ray sources
including variable and hard sources, along with pre-main sequence
stars and more evolved OB stars.
We summarize our findings here.

1. \chandra\ images show 304 X-ray sources; thirty percent of the sources
are hard, and two-thirds have near-infrared counterparts.  The full
list of \chandra\ X-ray sources is given in Table 1.

2. The multiple star system HD164492 is resolved for the first time  in
X-rays into individual components. X-ray emission is detected from
components A, B, C (a B6V star), which is blended with an 
unidentified source in the \chandra\ images.
This blend has 
comparable
X-ray brightness to the O star. The O star HD 164492A shows small but
significant variability and has a  soft spectrum with a temperature of
0.5 keV. The temperature is comparable to those of other single massive
stars and the  ratio of X-ray and bolometric luminosities  is smaller
than the canonical value $L_{x}/L_{bol} \approx -7$ but it is within
the scatter of distribution. The lack of any hard component suggests that
neither a magnetically confined wind shock nor colliding wind shock is
needed to describe  the X-ray emission from the O star.  The
variability of the X-ray  emission implies that the emission is
produced by a small number of  strong shocks in the wind of HD 164492A.

3. The X-ray spectrum from the component C blend requires a two-temperature
thermal model with $kT_1 = 0.6\pm0.4$ keV and $kT_2 = 5.9
(^{+\infty}_{-3.4})$ keV. The inferred X-ray luminosity is 2$\times
10^{32}$ erg s$^{-1}$.   This blend is highly variable in X-rays which
suggests that one of the stars dominates the emission.

4. We found a dozen stars which show evidence of flaring activity 
and there could be many as $\sim$ 40 variable stars in the full 
source list. Nine sources
(sources 8, 23, 97, 166, 194, 237, 256, 283 and 285 in Table 2)
have significant variability based  on the $\chi^2$
statistics. We searched for their near-infrared and optical
counterparts, and found that six stars have both optical and 2MASS
counterparts.  These sources are likely in later stages of PMS evolution.
Four sources which have no optical counterparts and have SHR greather than
$-0.2$ are likely in early stages of PMS, possibly Class I or
earlier.   There are a few stars with neither near-IR nor optical
counterpart whose light curves show strong flares, suggesting that they are 
very  early stage pre-main sequence stars.

5. We detected X-ray emission from  TC1 and possibly TC4,  two massive
star forming cores with  bipolar wings  and associated Class  0
candidates. Both TC1 and TC4  show extremely hard X-rays and their
spectra imply very high absorption  (N$_H$=3.5 - 6 $\times 10^{22}$
cm$^{-2}$) and  high luminosity ($2 - 5\times 10^{31}$ erg s$^{-1}$).
Only one other detection of X-ray emission from a  Class 0 object has been
previously reported. Thus our two detections imply the second and third
X-ray detection of a Class 0 object. It is unclear if the X-ray
emission of these objects is due to solar-type magnetic activities as
in Class I objects. If the X-ray emission is  from the
accreting stage, the X-ray  emission may be related
to the competing processes of accumulation of angular momentum toward
the growing central star and release of angular momentum by outflow
processes.

Thirty percent of the X-ray sources in the full field are shown to be 
hard (SHR $> -0.2$) sources
(shown in blue in Figure ~\ref{chandrazoom}), and
16 percent of sources are extremely hard sources (SHR $> +0.2$). A high
proportion of these sources are probably PMS because they are either 
highly embedded or  extemely high
temperature.
These hard sources  along
with the sources near HD 164492C  are responsible for
the hard spectra seen by \emph{ASCA}. 
The accurate positions from the  complete list of the
X-ray sources are provided in Table 1. 
These sources provide an opportunity
to identify interesting PMS  using the multi-wavelength follow-up
observations, which will help us to further understand the population
and evolution of protostellar objects.

\acknowledgements
Partial support for this work was provided by NASA through {\it Chandra}
grant G02-3095A.
J. R.  and S. V. R. acknowledge the support
of California Institute of Technology, the Jet Propulsion Laboratory,
which is operated under contract with NASA.

\begin{deluxetable}{rccrrr}
\tablenum{1}
\tablewidth{0pt}
\tablecaption{Catalog of X-ray sources}\label{catalog}
\tablehead{
\colhead{No} &
\colhead{WID} &
\colhead{CXOM20} &
\colhead{Net Counts} &
\colhead{SNR} &
\colhead{Hardness} \\
\colhead{} &
\colhead{} &
\colhead{} &
\colhead{cts ks$^{-1}$} &
\colhead{} &
\colhead{Ratio}
}
\startdata
  1 & 177 & 18:01:50.61--22:54:39.6 &  12 $\pm$  4 &   3 & -0.23 $\pm$ 0.20  \\
  2 & 208 & 18:01:50.79--22:52:18.5 &  22 $\pm$  7 &   3 &  0.09 $\pm$ 0.13  \\
  3 & 206 & 18:01:52.54--22:57:33.6 &  22 $\pm$  6 &   4 & -0.26 $\pm$ 0.16  \\
  4 & 207 & 18:01:53.57--22:51:35.6 &  24 $\pm$  7 &   4 &  0.56 $\pm$ 0.16  \\
  5 & 333 & 18:01:57.18--23:01:22.1 &  58 $\pm$  9 &  11 & -0.49 $\pm$ 0.14  \\
  6 & 344 & 18:01:57.64--23:00:01.7 &  13 $\pm$  4 &   3 & -0.19 $\pm$ 0.20  \\
  7 & 197 & 18:01:58.33--22:52:26.5 &  21 $\pm$  6 &   5 & -0.36 $\pm$ 0.14  \\
  8 & 196 & 18:01:59.07--22:52:27.5 &  73 $\pm$ 10 &  12 & -0.49 $\pm$ 0.11  \\
  9 & 352 & 18:01:59.12--23:00:55.3 &  20 $\pm$  5 &   5 &  0.76 $\pm$ 0.22  \\
 10 & 195 & 18:01:59.85--22:53:14.8 &  41 $\pm$  7 &   9 & -0.62 $\pm$ 0.16  \\
 11 & 203 & 18:02:02.04--22:54:19.1 &  14 $\pm$  4 &   4 & -0.36 $\pm$ 0.23  \\
 12 & 348 & 18:02:02.07--23:00:56.2 &  13 $\pm$  4 &   4 &  0.69 $\pm$ 0.10  \\
 13 & 332 & 18:02:02.54--23:02:32.5 & 103 $\pm$ 11 &  20 &  0.81 $\pm$ 0.12  \\
 14 & 343 & 18:02:03.13--22:59:09.6 &  13 $\pm$  4 &   4 & -0.33 $\pm$ 0.19  \\
 15 & 201 & 18:02:03.45--22:56:07.2 &  13 $\pm$  4 &   4 & -0.69 $\pm$ 0.34  \\
 16 & 194 & 18:02:03.86--22:52:58.0 &  45 $\pm$  7 &  10 &  0.84 $\pm$ 0.18  \\
 17 & 313 & 18:02:04.75--23:04:03.0 &  11 $\pm$  4 &   3 &  0.29 $\pm$ 0.19  \\
 18 & 189 & 18:02:05.50--22:55:01.2 &  19 $\pm$  5 &   6 & -0.65 $\pm$ 0.25  \\
 19 & 341 & 18:02:06.05--23:00:49.9 &  14 $\pm$  4 &   4 & -0.60 $\pm$ 0.30  \\
 20 & 175 & 18:02:07.02--22:57:25.8 &  13 $\pm$  4 &   6 & -0.53 $\pm$ 0.27  \\
 21 & 174 & 18:02:07.62--22:57:42.7 & 102 $\pm$ 10 &  37 & -0.51 $\pm$ 0.11  \\
 22 & 188 & 18:02:07.75--22:57:23.4 &  22 $\pm$  5 &   8 &  0.92 $\pm$ 0.28  \\
 23 & 173 & 18:02:07.77--22:55:33.9 &  58 $\pm$  8 &  19 & -0.58 $\pm$ 0.15  \\
 24 & 172 & 18:02:08.09--22:57:18.6 &  39 $\pm$  6 &  14 & -0.85 $\pm$ 0.21  \\
 25 & 289 & 18:02:08.52--22:59:00.0 &  62 $\pm$  8 &  24 & -0.76 $\pm$ 0.15  \\
 26 & 354 & 18:02:09.18--23:02:36.9 &   5 $\pm$  2 &   2 &  0.00 $\pm$ 0.22  \\
 27 & 205 & 18:02:09.42--22:50:56.1 &  10 $\pm$  4 &   3 & -0.12 $\pm$ 0.20  \\
 28 & 339 & 18:02:09.92--23:04:27.5 &  43 $\pm$  8 &   8 & -0.47 $\pm$ 0.15  \\
 29 & 186 & 18:02:10.46--22:55:13.8 &  28 $\pm$  5 &  11 & -0.66 $\pm$ 0.22  \\
 30 & 185 & 18:02:10.80--22:55:32.6 &  14 $\pm$  4 &   6 & -0.25 $\pm$ 0.26  \\
 31 & 353 & 18:02:11.27--23:04:11.9 &  18 $\pm$  5 &   4 &  0.14 $\pm$ 0.17  \\
 32 & 160 & 18:02:11.36--22:56:50.8 &  74 $\pm$  9 &  29 & -0.53 $\pm$ 0.13  \\
 33 & 246 & 18:02:11.63--22:59:33.9 & 189 $\pm$ 14 &  70 & -0.54 $\pm$ 0.08  \\
 34 & 204 & 18:02:11.66--22:50:10.8 &  21 $\pm$  6 &   4 &  0.31 $\pm$ 0.19  \\
 35 & 170 & 18:02:11.78--22:56:45.8 &  34 $\pm$  6 &  14 &  0.24 $\pm$ 0.18  \\
 36 & 159 & 18:02:11.95--22:55:43.6 & 294 $\pm$ 17 &  92 & -0.70 $\pm$ 0.07  \\
 37 & 351 & 18:02:12.26--23:03:10.4 &  14 $\pm$  4 &   3 &  0.33 $\pm$ 0.22  \\
 38 & 347 & 18:02:12.47--23:05:48.1 &  62 $\pm$  9 &   9 &  0.67 $\pm$ 0.13  \\
 39 & 311 & 18:02:12.68--22:58:52.2 &   6 $\pm$  2 &   2 &  1.00 $\pm$ 0.45  \\
 40 & 199 & 18:02:12.79--22:57:00.0 &   6 $\pm$  3 &   3 &  0.80 $\pm$ 0.40  \\
 41 & 288 & 18:02:13.15--23:01:59.3 &  92 $\pm$ 10 &  27 & -0.66 $\pm$ 0.12  \\
 42 & 338 & 18:02:13.38--23:04:13.5 &   9 $\pm$  4 &   3 &  0.38 $\pm$ 0.27  \\
 43 & 192 & 18:02:13.50--22:57:31.4 &   6 $\pm$  2 &   2 &  0.00 $\pm$ 0.35  \\
 44 & 287 & 18:02:13.83--22:58:36.1 &  30 $\pm$  6 &  13 & -0.69 $\pm$ 0.21  \\
 45 & 286 & 18:02:13.95--23:02:12.7 & 115 $\pm$ 11 &  29 & -0.23 $\pm$ 0.09  \\
 46 & 168 & 18:02:14.37--22:54:21.5 &  55 $\pm$  7 &  22 & -0.75 $\pm$ 0.17  \\
 47 & 285 & 18:02:15.56--23:02:22.5 &  47 $\pm$  7 &  14 & -0.49 $\pm$ 0.15  \\
 48 & 350 & 18:02:15.79--23:05:56.1 &  20 $\pm$  6 &   4 & -0.43 $\pm$ 0.14  \\
 49 & 284 & 18:02:15.85--22:58:27.4 &  14 $\pm$  4 &   6 & -0.60 $\pm$ 0.30  \\
 50 & 167 & 18:02:16.17--22:57:23.4 &   9 $\pm$  3 &   4 & -1.00 $\pm$ 0.47  \\
 51 & 244 & 18:02:16.44--22:59:39.5 &  36 $\pm$  6 &  15 & -0.78 $\pm$ 0.21  \\
 52 & 283 & 18:02:16.61--22:58:36.8 &  11 $\pm$  3 &   5 & -0.67 $\pm$ 0.35  \\
 53 & 282 & 18:02:16.81--23:03:47.0 & 167 $\pm$ 14 &  31 & -0.63 $\pm$ 0.09  \\
 54 & 280 & 18:02:17.01--22:59:17.8 &  13 $\pm$  4 &   6 & -1.00 $\pm$ 0.41  \\
 55 & 279 & 18:02:17.28--22:57:50.9 &  22 $\pm$  5 &  10 &  0.91 $\pm$ 0.28  \\
 56 & 202 & 18:02:17.36--22:53:44.6 &   7 $\pm$  3 &   3 & -0.56 $\pm$ 0.38  \\
 57 & 156 & 18:02:17.65--22:56:18.0 &  26 $\pm$  5 &  12 & -0.85 $\pm$ 0.26  \\
 58 & 310 & 18:02:17.91--23:04:07.3 &  10 $\pm$  4 &   3 & -0.67 $\pm$ 0.35  \\
 59 & 309 & 18:02:17.97--23:00:14.0 &   9 $\pm$  3 &   4 & -0.82 $\pm$ 0.39  \\
 60 & 331 & 18:02:18.18--23:03:43.4 &  30 $\pm$  6 &   8 & -0.89 $\pm$ 0.22  \\
 61 & 337 & 18:02:18.78--23:03:42.4 &  37 $\pm$  7 &  10 & -0.57 $\pm$ 0.17  \\
 62 & 198 & 18:02:18.90--22:51:45.1 &  26 $\pm$  6 &   7 & -0.66 $\pm$ 0.20  \\
 63 & 336 & 18:02:19.09--23:03:32.7 &  11 $\pm$  4 &   3 &  0.00 $\pm$ 0.24  \\
 64 & 155 & 18:02:19.17--22:53:46.3 &  97 $\pm$ 10 &  34 & -0.77 $\pm$ 0.13  \\
 65 & 243 & 18:02:19.19--22:58:36.7 &  15 $\pm$  4 &   6 & -0.60 $\pm$ 0.30  \\
 66 & 242 & 18:02:19.70--22:59:40.6 &  11 $\pm$  3 &   5 & -1.00 $\pm$ 0.45  \\
 67 & 278 & 18:02:19.98--23:01:21.3 &  17 $\pm$  4 &   7 & -0.14 $\pm$ 0.22  \\
 68 & 330 & 18:02:20.43--23:03:15.7 &  18 $\pm$  5 &   5 &  0.68 $\pm$ 0.24  \\
 69 & 165 & 18:02:20.57--22:53:56.2 &  18 $\pm$  4 &   7 & -0.37 $\pm$ 0.24  \\
 70 & 346 & 18:02:20.76--23:02:41.1 &  15 $\pm$  4 &   4 & -0.08 $\pm$ 0.20  \\
 71 & 329 & 18:02:21.05--22:59:02.5 &   6 $\pm$  2 &   2 & -1.00 $\pm$ 0.53  \\
 72 & 308 & 18:02:21.06--23:01:43.6 &  40 $\pm$  7 &  12 & -0.45 $\pm$ 0.17  \\
 73 & 241 & 18:02:21.07--23:03:13.9 & 507 $\pm$ 23 &  97 & -0.51 $\pm$ 0.05  \\
 74 & 328 & 18:02:21.20--23:00:17.3 &   3 $\pm$  2 &   1 &  0.33 $\pm$ 0.61  \\
 75 & 240 & 18:02:21.26--22:57:50.1 &  54 $\pm$  7 &  21 & -0.53 $\pm$ 0.16  \\
 76 & 335 & 18:02:21.27--23:01:28.8 &   9 $\pm$  3 &   3 & -0.57 $\pm$ 0.31  \\
 77 & 239 & 18:02:21.46--23:01:02.5 &  66 $\pm$  8 &  23 & -0.55 $\pm$ 0.14  \\
 78 & 238 & 18:02:21.93--23:00:57.7 & 114 $\pm$ 11 &  39 & -0.72 $\pm$ 0.12  \\
 79 & 307 & 18:02:21.96--23:02:07.2 &  31 $\pm$  6 &   8 & -0.38 $\pm$ 0.17  \\
 80 & 190 & 18:02:22.04--22:54:15.7 &   9 $\pm$  3 &   4 & -0.64 $\pm$ 0.36  \\
 81 & 277 & 18:02:22.24--23:01:50.5 &  38 $\pm$  7 &   8 & -0.44 $\pm$ 0.15  \\
 82 & 306 & 18:02:22.27--23:00:46.4 &  14 $\pm$  4 &   6 & -0.76 $\pm$ 0.31  \\
 83 & 049 & 18:02:22.36--22:49:09.3 &  41 $\pm$  8 &   7 &  0.68 $\pm$ 0.16  \\
 84 & 179 & 18:02:22.36--22:56:35.2 &   9 $\pm$  3 &   4 &  0.20 $\pm$ 0.32  \\
 85 & 276 & 18:02:22.42--23:02:10.7 &  80 $\pm$  9 &  18 & -0.67 $\pm$ 0.13  \\
 86 & 275 & 18:02:22.61--23:01:58.9 & 116 $\pm$ 12 &  20 & -0.41 $\pm$ 0.09  \\
 87 & 305 & 18:02:22.78--23:00:18.5 &   7 $\pm$  3 &   3 &  1.00 $\pm$ 0.45  \\
 88 & 327 & 18:02:22.81--23:01:52.3 &   5 $\pm$  3 &   1 & -0.67 $\pm$ 0.25  \\
 89 & 237 & 18:02:22.86--22:59:35.6 &  29 $\pm$  5 &  12 & -0.03 $\pm$ 0.19  \\
 90 & 274 & 18:02:22.90--23:01:28.5 &  35 $\pm$  6 &  10 & -0.12 $\pm$ 0.15  \\
 91 & 235 & 18:02:23.05--23:01:58.1 & 857 $\pm$ 30 & 129 & -0.47 $\pm$ 0.03  \\
 92 & 273 & 18:02:23.06--23:01:25.3 &   8 $\pm$  3 &   3 & -0.64 $\pm$ 0.36  \\
 93 & 236 & 18:02:23.09--23:01:44.8 & 317 $\pm$ 18 &  52 & -0.63 $\pm$ 0.06  \\
 94 & 234 & 18:02:23.16--23:02:00.3 & 831 $\pm$ 29 & 119 & -0.50 $\pm$ 0.03  \\
 95 & 304 & 18:02:23.19--23:00:23.6 &   9 $\pm$  3 &   4 & -0.69 $\pm$ 0.34  \\
 96 & 233 & 18:02:23.21--23:00:12.3 &  17 $\pm$  4 &   7 & -0.58 $\pm$ 0.27  \\
 97 & 272 & 18:02:23.25--23:01:35.0 &  48 $\pm$  7 &  12 & -0.18 $\pm$ 0.13  \\
 98 & 303 & 18:02:23.32--23:03:11.3 &  20 $\pm$  5 &   6 &  0.00 $\pm$ 0.20  \\
 99 & 232 & 18:02:23.41--23:01:41.5 &  90 $\pm$ 10 &  20 & -0.75 $\pm$ 0.11  \\
100 & 326 & 18:02:23.41--23:02:22.4 &  19 $\pm$  5 &   5 & -0.71 $\pm$ 0.23  \\
101 & 231 & 18:02:23.51--23:00:23.3 &  65 $\pm$  8 &  25 & -0.13 $\pm$ 0.12  \\
102 & 230 & 18:02:23.54--23:01:50.9 & 884 $\pm$ 30 & 114 & -0.89 $\pm$ 0.04  \\
103 & 270 & 18:02:23.63--23:03:27.7 & 155 $\pm$ 13 &  33 & -0.18 $\pm$ 0.08  \\
104 & 268 & 18:02:23.65--23:00:55.6 &  40 $\pm$  6 &  15 & -0.29 $\pm$ 0.16  \\
105 & 269 & 18:02:23.66--23:01:57.3 &  35 $\pm$  7 &   8 & -0.60 $\pm$ 0.16  \\
106 & 229 & 18:02:23.71--23:01:45.4 & 299 $\pm$ 18 &  47 & -0.64 $\pm$ 0.07  \\
107 & 267 & 18:02:23.83--22:59:39.0 &  21 $\pm$  5 &   9 & -0.45 $\pm$ 0.23  \\
108 & 164 & 18:02:23.95--22:54:30.7 &  23 $\pm$  5 &  10 & -0.67 $\pm$ 0.22  \\
109 & 302 & 18:02:23.98--23:04:19.2 & 204 $\pm$ 15 &  35 & -0.51 $\pm$ 0.08  \\
110 & 301 & 18:02:24.17--23:00:36.6 &   7 $\pm$  3 &   3 & -0.80 $\pm$ 0.40  \\
111 & 020 & 18:02:24.19--22:51:13.2 &  12 $\pm$  4 &   4 & -0.87 $\pm$ 0.34  \\
112 & 228 & 18:02:24.29--22:59:27.1 &  72 $\pm$  9 &  30 & -0.75 $\pm$ 0.15  \\
113 & 266 & 18:02:24.57--23:00:20.7 &  20 $\pm$  5 &   8 & -0.24 $\pm$ 0.22  \\
114 & 227 & 18:02:24.63--23:01:03.7 &  76 $\pm$  9 &  26 &  0.26 $\pm$ 0.12  \\
115 & 154 & 18:02:24.90--22:55:26.4 &  18 $\pm$  4 &   8 & -0.89 $\pm$ 0.31  \\
116 & 265 & 18:02:25.12--23:02:17.4 &  79 $\pm$  9 &  23 & -0.76 $\pm$ 0.14  \\
117 & 264 & 18:02:25.16--23:01:26.7 &  41 $\pm$  7 &  15 &  0.24 $\pm$ 0.16  \\
118 & 010 & 18:02:25.17--22:53:41.0 &  14 $\pm$  4 &   6 & -0.47 $\pm$ 0.28  \\
119 & 323 & 18:02:25.34--23:03:55.5 &  83 $\pm$ 10 &  17 & -0.56 $\pm$ 0.13  \\
120 & 324 & 18:02:25.35--23:02:56.7 &  32 $\pm$  6 &   9 & -0.57 $\pm$ 0.19  \\
121 & 226 & 18:02:25.37--23:00:36.3 &  25 $\pm$  5 &  10 & -0.77 $\pm$ 0.25  \\
122 & 300 & 18:02:25.43--23:01:39.9 &  27 $\pm$  5 &   9 & -0.47 $\pm$ 0.20  \\
123 & 262 & 18:02:25.63--23:00:56.9 &  50 $\pm$  7 &  19 & -0.37 $\pm$ 0.15  \\
124 & 261 & 18:02:25.84--23:01:42.6 &  25 $\pm$  5 &   9 & -0.61 $\pm$ 0.21  \\
125 & 225 & 18:02:25.93--23:00:11.6 &  69 $\pm$  8 &  28 & -0.67 $\pm$ 0.10  \\
126 & 224 & 18:02:25.94--23:00:34.6 &  87 $\pm$  9 &  33 & -0.57 $\pm$ 0.12  \\
127 & 260 & 18:02:25.95--22:58:03.2 &  16 $\pm$  4 &   7 &  0.05 $\pm$ 0.23  \\
128 & 223 & 18:02:25.99--23:00:13.4 & 170 $\pm$ 13 &  62 & -0.77 $\pm$ 0.09  \\
129 & 178 & 18:02:25.99--22:55:27.8 &   4 $\pm$  2 &   2 & -0.20 $\pm$ 0.46  \\
130 & 259 & 18:02:26.01--23:02:27.7 &  61 $\pm$  8 &  19 & -0.60 $\pm$ 0.14  \\
131 & 258 & 18:02:26.09--23:01:28.6 &   5 $\pm$  2 &   2 & -1.00 $\pm$ 0.63  \\
132 & 321 & 18:02:26.47--23:01:55.9 &  16 $\pm$  4 &   5 & -0.40 $\pm$ 0.24  \\
133 & 222 & 18:02:26.85--23:00:14.3 &  61 $\pm$  8 &  24 & -0.48 $\pm$ 0.14  \\
134 & 019 & 18:02:26.95--22:51:07.9 & 108 $\pm$ 11 &  25 & -0.68 $\pm$ 0.11  \\
135 & 256 & 18:02:27.13--23:00:50.6 &   9 $\pm$  3 &   4 & -0.09 $\pm$ 0.30  \\
136 & 255 & 18:02:27.17--23:03:33.4 & 215 $\pm$ 15 &  45 & -0.75 $\pm$ 0.08  \\
137 & 220 & 18:02:27.32--22:59:37.7 &  34 $\pm$  6 &  14 & -0.71 $\pm$ 0.21  \\
138 & 349 & 18:02:27.45--23:04:13.7 &  17 $\pm$  5 &   4 & -0.56 $\pm$ 0.20  \\
139 & 299 & 18:02:27.56--23:00:20.3 &   8 $\pm$  3 &   3 & -0.56 $\pm$ 0.38  \\
140 & 254 & 18:02:27.97--22:58:56.9 &  11 $\pm$  3 &   5 & -0.82 $\pm$ 0.39  \\
141 & 298 & 18:02:28.03--23:01:43.0 &  19 $\pm$  5 &   6 & -0.39 $\pm$ 0.22  \\
142 & 320 & 18:02:28.13--23:02:11.1 &  11 $\pm$  4 &   4 & -0.75 $\pm$ 0.31  \\
143 & 253 & 18:02:28.41--22:59:47.3 &  25 $\pm$  5 &  10 & -0.54 $\pm$ 0.22  \\
144 & 004 & 18:02:28.44--22:55:44.5 &  21 $\pm$  5 &   9 & -1.00 $\pm$ 0.32  \\
145 & 319 & 18:02:28.44--23:01:14.1 &  11 $\pm$  3 &   4 & -0.69 $\pm$ 0.34  \\
146 & 219 & 18:02:28.51--23:00:37.7 & 297 $\pm$ 17 &  95 & -0.36 $\pm$ 0.06  \\
147 & 297 & 18:02:28.52--23:00:14.6 &   7 $\pm$  3 &   3 &  1.00 $\pm$ 0.45  \\
148 & 048 & 18:02:28.54--22:51:45.8 &  13 $\pm$  4 &   4 & -0.39 $\pm$ 0.22  \\
149 & 218 & 18:02:28.65--23:03:03.4 & 432 $\pm$ 21 &  86 & -0.53 $\pm$ 0.05  \\
150 & 318 & 18:02:28.66--23:01:36.8 &   6 $\pm$  3 &   2 & -0.25 $\pm$ 0.26  \\
151 & 217 & 18:02:28.74--22:59:47.4 & 269 $\pm$ 16 &  91 & -0.62 $\pm$ 0.07  \\
152 & 216 & 18:02:28.80--22:59:00.1 &  54 $\pm$  7 &  22 & -0.41 $\pm$ 0.15  \\
153 & 317 & 18:02:28.81--23:02:25.5 &  21 $\pm$  5 &   7 & -0.89 $\pm$ 0.22  \\
154 & 252 & 18:02:29.10--23:00:40.9 &  19 $\pm$  4 &   8 &  0.20 $\pm$ 0.23  \\
155 & 009 & 18:02:29.14--22:52:52.6 & 108 $\pm$ 11 &  32 & -0.75 $\pm$ 0.12  \\
156 & 215 & 18:02:29.35--23:01:01.6 &  29 $\pm$  5 &  11 & -0.47 $\pm$ 0.20  \\
157 & 214 & 18:02:29.47--22:59:14.7 &  25 $\pm$  5 &  11 & -0.56 $\pm$ 0.22  \\
158 & 018 & 18:02:29.53--22:55:36.5 &  13 $\pm$  4 &   5 & -0.73 $\pm$ 0.32  \\
159 & 117 & 18:02:29.59--22:56:54.4 &   4 $\pm$  2 &   2 & -1.00 $\pm$ 0.71  \\
160 & 213 & 18:02:29.60--22:58:59.3 &  16 $\pm$  4 &   7 & -0.50 $\pm$ 0.28  \\
161 & 040 & 18:02:29.68--22:50:12.5 &  14 $\pm$  4 &   4 &  0.90 $\pm$ 0.29  \\
162 & 003 & 18:02:29.79--22:56:03.5 &  24 $\pm$  5 &  11 & -0.67 $\pm$ 0.25  \\
163 & 345 & 18:02:30.03--23:02:02.5 &   4 $\pm$  2 &   1 &  1.00 $\pm$ 0.39  \\
164 & 250 & 18:02:30.19--23:00:39.5 &  30 $\pm$  6 &  12 &  0.44 $\pm$ 0.19  \\
165 & 251 & 18:02:30.20--22:59:50.2 &  10 $\pm$  3 &   4 & -0.45 $\pm$ 0.33  \\
166 & 212 & 18:02:30.37--22:59:29.1 & 135 $\pm$ 12 &  57 & -0.19 $\pm$ 0.09  \\
167 & 249 & 18:02:30.38--23:01:34.5 &  96 $\pm$ 10 &  30 & -0.74 $\pm$ 0.12  \\
168 & 053 & 18:02:30.47--22:48:58.6 &  40 $\pm$  8 &   7 &  0.47 $\pm$ 0.14  \\
169 & 032 & 18:02:30.89--22:53:03.8 &  16 $\pm$  4 &   5 & -0.79 $\pm$ 0.29  \\
170 & 211 & 18:02:30.93--23:02:35.6 &  73 $\pm$  9 &  21 & -0.79 $\pm$ 0.15  \\
171 & 210 & 18:02:30.98--23:00:43.6 & 128 $\pm$ 11 &  43 &  0.48 $\pm$ 0.10  \\
172 & 316 & 18:02:31.10--23:00:31.5 &   9 $\pm$  3 &   4 &  0.33 $\pm$ 0.30  \\
173 & 296 & 18:02:31.20--23:01:24.4 &  36 $\pm$  6 &  13 & -0.45 $\pm$ 0.17  \\
174 & 295 & 18:02:31.22--23:01:20.8 &  18 $\pm$  4 &   7 & -0.50 $\pm$ 0.21  \\
175 & 294 & 18:02:31.55--23:01:49.4 &  28 $\pm$  6 &   9 & -0.71 $\pm$ 0.23  \\
176 & 017 & 18:02:31.67--22:55:12.1 &  10 $\pm$  3 &   4 &  0.45 $\pm$ 0.33  \\
177 & 209 & 18:02:31.71--23:02:16.8 &  97 $\pm$ 10 &  30 & -0.59 $\pm$ 0.11  \\
178 & 315 & 18:02:31.82--23:03:33.9 &  13 $\pm$  4 &   4 & -0.65 $\pm$ 0.29  \\
179 & 293 & 18:02:31.84--23:00:24.9 &   8 $\pm$  3 &   3 & -0.85 $\pm$ 0.36  \\
180 & 096 & 18:02:31.89--22:58:02.0 &   7 $\pm$  3 &   3 &  0.43 $\pm$ 0.41  \\
181 & 140 & 18:02:31.94--23:00:24.7 &   5 $\pm$  2 &   2 & -0.83 $\pm$ 0.38  \\
182 & 248 & 18:02:31.99--23:02:40.4 & 164 $\pm$ 13 &  38 & -0.79 $\pm$ 0.10  \\
183 & 016 & 18:02:32.00--22:53:59.8 &  15 $\pm$  4 &   6 & -0.18 $\pm$ 0.25  \\
184 & 095 & 18:02:32.30--22:59:27.9 &   8 $\pm$  3 &   3 & -0.75 $\pm$ 0.44  \\
185 & 314 & 18:02:32.38--23:04:54.8 &  40 $\pm$  7 &   8 & -0.42 $\pm$ 0.15  \\
186 & 008 & 18:02:32.40--22:56:16.0 &  14 $\pm$  4 &   6 &  0.86 $\pm$ 0.35  \\
187 & 015 & 18:02:32.60--22:53:25.6 &   8 $\pm$  3 &   4 & -0.67 $\pm$ 0.35  \\
188 & 292 & 18:02:32.61--23:01:16.4 &  21 $\pm$  5 &   9 & -0.82 $\pm$ 0.28  \\
189 & 094 & 18:02:32.61--22:59:24.5 &  15 $\pm$  4 &   7 & -0.75 $\pm$ 0.31  \\
190 & 132 & 18:02:32.63--22:59:37.6 &  10 $\pm$  3 &   4 & -0.27 $\pm$ 0.31  \\
191 & 093 & 18:02:32.67--22:59:49.5 &  27 $\pm$  5 &  11 & -0.50 $\pm$ 0.21  \\
192 & 116 & 18:02:32.68--22:59:46.3 &  14 $\pm$  4 &   6 &  0.12 $\pm$ 0.25  \\
193 & 115 & 18:02:32.74--23:00:46.7 &  16 $\pm$  4 &   7 & -0.18 $\pm$ 0.25  \\
194 & 291 & 18:02:32.75--23:04:21.0 &  74 $\pm$  9 &  14 & -0.58 $\pm$ 0.13  \\
195 & 114 & 18:02:32.79--22:59:20.5 &   8 $\pm$  3 &   3 &  1.00 $\pm$ 0.45  \\
196 & 290 & 18:02:33.02--23:03:10.8 &  73 $\pm$  9 &  18 &  0.83 $\pm$ 0.14  \\
197 & 002 & 18:02:33.03--22:55:47.7 &  22 $\pm$  5 &   9 &  0.13 $\pm$ 0.21  \\
198 & 334 & 18:02:33.16--23:02:37.4 &   9 $\pm$  3 &   3 & -0.62 $\pm$ 0.29  \\
199 & 092 & 18:02:33.34--23:01:11.3 &   9 $\pm$  3 &   4 & -0.54 $\pm$ 0.32  \\
200 & 091 & 18:02:33.53--23:01:17.4 &  34 $\pm$  6 &  13 & -0.63 $\pm$ 0.19  \\
201 & 113 & 18:02:33.67--22:59:54.3 &   9 $\pm$  3 &   4 & -1.00 $\pm$ 0.43  \\
202 & 039 & 18:02:33.67--22:52:57.4 &  10 $\pm$  3 &   3 &  1.00 $\pm$ 0.41  \\
203 & 001 & 18:02:33.78--22:55:37.5 &  99 $\pm$ 10 &  40 & -0.72 $\pm$ 0.12  \\
204 & 074 & 18:02:33.78--22:58:40.9 &  24 $\pm$  5 &  11 & -0.92 $\pm$ 0.27  \\
205 & 031 & 18:02:33.90--22:53:22.8 &  18 $\pm$  4 &   7 & -0.57 $\pm$ 0.22  \\
206 & 090 & 18:02:33.91--22:59:52.6 &  25 $\pm$  5 &  10 & -0.04 $\pm$ 0.20  \\
207 & 030 & 18:02:34.02--22:50:13.4 &  77 $\pm$ 10 &  15 & -0.71 $\pm$ 0.13  \\
208 & 247 & 18:02:34.03--23:03:17.9 & 182 $\pm$ 14 &  36 & -0.50 $\pm$ 0.08  \\
209 & 073 & 18:02:34.33--23:01:15.4 & 154 $\pm$ 13 &  47 & -0.66 $\pm$ 0.10  \\
210 & 112 & 18:02:34.45--22:58:10.8 &  11 $\pm$  3 &   5 & -0.23 $\pm$ 0.28  \\
211 & 072 & 18:02:34.51--22:59:38.0 & 312 $\pm$ 18 & 107 & -0.47 $\pm$ 0.06  \\
212 & 071 & 18:02:34.57--23:00:30.9 &  45 $\pm$  7 &  16 & -0.32 $\pm$ 0.16  \\
213 & 047 & 18:02:34.70--22:51:25.3 &  10 $\pm$  4 &   3 &  0.14 $\pm$ 0.27  \\
214 & 111 & 18:02:34.85--23:00:46.1 &   9 $\pm$  3 &   4 &  1.00 $\pm$ 0.35  \\
215 & 131 & 18:02:34.91--23:01:48.7 &  16 $\pm$  4 &   5 &  0.14 $\pm$ 0.22  \\
216 & 089 & 18:02:35.05--23:00:52.6 &  79 $\pm$  9 &  28 & -0.68 $\pm$ 0.13  \\
217 & 110 & 18:02:35.24--23:01:02.3 &  10 $\pm$  3 &   4 & -0.20 $\pm$ 0.26  \\
218 & 130 & 18:02:35.33--23:00:09.7 &   8 $\pm$  3 &   3 & -0.33 $\pm$ 0.35  \\
219 & 070 & 18:02:35.55--22:59:55.5 &  92 $\pm$ 10 &  36 & -0.74 $\pm$ 0.13  \\
220 & 109 & 18:02:35.74--23:02:58.6 & 145 $\pm$ 12 &  35 & -0.41 $\pm$ 0.09  \\
221 & 088 & 18:02:35.95--22:59:47.0 &  21 $\pm$  5 &   8 & -0.70 $\pm$ 0.24  \\
222 & 068 & 18:02:35.98--23:01:40.8 & 182 $\pm$ 14 &  52 & -0.55 $\pm$ 0.08  \\
223 & 057 & 18:02:36.07--22:49:39.5 &  14 $\pm$  5 &   3 &  0.11 $\pm$ 0.17  \\
224 & 108 & 18:02:36.09--22:59:19.9 &   8 $\pm$  3 &   3 & -0.11 $\pm$ 0.34  \\
225 & 087 & 18:02:36.11--23:01:31.9 &  60 $\pm$  8 &  19 & -0.78 $\pm$ 0.16  \\
226 & 107 & 18:02:36.29--22:57:33.3 &   9 $\pm$  3 &   4 & -0.64 $\pm$ 0.36  \\
227 & 067 & 18:02:36.45--22:59:59.6 &  24 $\pm$  5 &   9 & -0.76 $\pm$ 0.25  \\
228 & 086 & 18:02:36.70--22:58:04.2 &  32 $\pm$  6 &  13 & -0.88 $\pm$ 0.23  \\
229 & 066 & 18:02:36.81--23:00:24.0 &  48 $\pm$  7 &  18 & -0.65 $\pm$ 0.17  \\
230 & 139 & 18:02:36.83--23:04:00.6 &  19 $\pm$  5 &   6 & -0.31 $\pm$ 0.18  \\
231 & 151 & 18:02:36.92--23:04:14.8 &  16 $\pm$  5 &   4 & -0.48 $\pm$ 0.23  \\
232 & 046 & 18:02:37.27--22:47:42.9 &  30 $\pm$  7 &   5 & -0.54 $\pm$ 0.13  \\
233 & 129 & 18:02:37.33--23:01:04.1 &  11 $\pm$  4 &   4 & -0.69 $\pm$ 0.34  \\
234 & 106 & 18:02:37.44--23:00:48.7 &   9 $\pm$  3 &   4 & -0.71 $\pm$ 0.33  \\
235 & 128 & 18:02:37.76--23:02:11.9 &  10 $\pm$  3 &   3 & -0.85 $\pm$ 0.36  \\
236 & 065 & 18:02:38.26--22:58:54.8 & 322 $\pm$ 18 & 121 & -0.75 $\pm$ 0.07  \\
237 & 029 & 18:02:38.85--22:50:06.1 &  81 $\pm$ 10 &  13 &  0.24 $\pm$ 0.09  \\
238 & 045 & 18:02:38.94--22:50:21.0 &  22 $\pm$  6 &   5 &  0.72 $\pm$ 0.21  \\
239 & 127 & 18:02:39.03--22:59:50.5 &  13 $\pm$  4 &   5 & -0.20 $\pm$ 0.26  \\
240 & 055 & 18:02:39.07--22:48:22.8 &  43 $\pm$  9 &   6 & -0.03 $\pm$ 0.11  \\
241 & 104 & 18:02:39.59--23:03:44.0 &  87 $\pm$ 10 &  17 & -0.53 $\pm$ 0.11  \\
242 & 064 & 18:02:39.94--23:00:12.9 & 119 $\pm$ 11 &  43 & -0.58 $\pm$ 0.11  \\
243 & 103 & 18:02:40.11--23:01:45.8 &  23 $\pm$  5 &   7 & -0.70 $\pm$ 0.24  \\
244 & 084 & 18:02:40.18--23:04:04.5 &  48 $\pm$  8 &  11 &  0.84 $\pm$ 0.17  \\
245 & 054 & 18:02:40.21--22:48:34.8 &  26 $\pm$  7 &   4 & -0.05 $\pm$ 0.11  \\
246 & 063 & 18:02:40.42--22:57:53.3 &  58 $\pm$  8 &  23 & -0.86 $\pm$ 0.17  \\
247 & 083 & 18:02:40.61--23:01:28.4 &  85 $\pm$  9 &  26 & -0.52 $\pm$ 0.12  \\
248 & 014 & 18:02:40.66--22:55:17.5 &  14 $\pm$  4 &   6 & -0.71 $\pm$ 0.33  \\
249 & 138 & 18:02:40.83--23:00:54.6 &   6 $\pm$  3 &   2 &  0.45 $\pm$ 0.33  \\
250 & 052 & 18:02:41.25--22:49:05.4 &  46 $\pm$  8 &   7 & -0.17 $\pm$ 0.09  \\
251 & 082 & 18:02:41.25--23:01:14.0 & 110 $\pm$ 11 &  31 & -0.61 $\pm$ 0.11  \\
252 & 062 & 18:02:41.37--22:57:42.4 &  72 $\pm$  9 &  24 &  0.51 $\pm$ 0.13  \\
253 & 126 & 18:02:41.50--22:58:20.8 &   5 $\pm$  2 &   2 &  0.33 $\pm$ 0.43  \\
254 & 137 & 18:02:41.91--23:00:57.9 &   9 $\pm$  3 &   3 & -0.67 $\pm$ 0.28  \\
255 & 061 & 18:02:42.19--22:57:52.1 &  30 $\pm$  6 &  12 & -0.87 $\pm$ 0.24  \\
256 & 125 & 18:02:42.36--23:04:35.8 &  72 $\pm$  9 &  13 & -0.17 $\pm$ 0.11  \\
257 & 060 & 18:02:42.82--22:58:35.9 & 109 $\pm$ 11 &  40 & -0.54 $\pm$ 0.11  \\
258 & 100 & 18:02:42.95--23:03:12.4 & 123 $\pm$ 12 &  24 & -0.59 $\pm$ 0.10  \\
259 & 028 & 18:02:43.03--22:55:38.5 &  17 $\pm$  4 &   6 &  0.24 $\pm$ 0.22  \\
260 & 027 & 18:02:43.20--22:54:35.4 &  22 $\pm$  5 &   7 & -0.83 $\pm$ 0.27  \\
261 & 099 & 18:02:43.51--23:01:35.4 &  58 $\pm$  8 &  15 & -0.06 $\pm$ 0.13  \\
262 & 080 & 18:02:43.54--22:57:36.0 &  12 $\pm$  4 &   5 & -0.65 $\pm$ 0.29  \\
263 & 124 & 18:02:44.04--22:57:15.6 &   5 $\pm$  2 &   2 & -0.50 $\pm$ 0.40  \\
264 & 123 & 18:02:44.18--23:02:45.1 &  50 $\pm$  8 &  11 & -0.51 $\pm$ 0.14  \\
265 & 145 & 18:02:44.76--22:58:15.4 &   8 $\pm$  3 &   3 &  0.69 $\pm$ 0.34  \\
266 & 079 & 18:02:44.78--22:59:51.9 &  47 $\pm$  7 &  15 & -0.37 $\pm$ 0.15  \\
267 & 078 & 18:02:45.45--22:59:18.4 &  48 $\pm$  7 &  14 & -0.67 $\pm$ 0.17  \\
268 & 136 & 18:02:45.46--22:58:37.4 &   9 $\pm$  3 &   3 &  0.69 $\pm$ 0.34  \\
269 & 144 & 18:02:45.74--22:59:44.2 &   9 $\pm$  3 &   3 & -0.38 $\pm$ 0.30  \\
270 & 135 & 18:02:45.99--22:57:19.6 &   9 $\pm$  3 &   3 &  0.86 $\pm$ 0.35  \\
271 & 026 & 18:02:46.63--22:50:32.2 &  86 $\pm$ 11 &  13 & -0.80 $\pm$ 0.12  \\
272 & 051 & 18:02:46.95--22:53:22.5 &  10 $\pm$  4 &   3 & -0.09 $\pm$ 0.21  \\
273 & 037 & 18:02:47.20--22:55:02.4 &  11 $\pm$  4 &   4 & -0.44 $\pm$ 0.26  \\
274 & 044 & 18:02:47.49--22:48:20.4 &  42 $\pm$  9 &   6 & -0.20 $\pm$ 0.10  \\
275 & 121 & 18:02:47.72--22:58:39.3 &  23 $\pm$  5 &   8 & -0.85 $\pm$ 0.25  \\
276 & 120 & 18:02:47.82--22:59:42.2 &  22 $\pm$  5 &   6 &  0.33 $\pm$ 0.22  \\
277 & 036 & 18:02:47.90--22:49:11.5 &  57 $\pm$  9 &   8 & -0.61 $\pm$ 0.11  \\
278 & 119 & 18:02:48.38--23:00:30.6 &  40 $\pm$  7 &  12 & -0.27 $\pm$ 0.14  \\
279 & 077 & 18:02:48.63--22:59:32.4 &  73 $\pm$  9 &  18 & -0.82 $\pm$ 0.15  \\
280 & 035 & 18:02:48.70--22:54:17.6 &  20 $\pm$  5 &   6 & -0.67 $\pm$ 0.25  \\
281 & 043 & 18:02:48.75--22:51:57.1 &  13 $\pm$  4 &   3 & -0.58 $\pm$ 0.24  \\
282 & 143 & 18:02:49.02--22:56:57.4 &   8 $\pm$  3 &   3 & -0.68 $\pm$ 0.28  \\
283 & 076 & 18:02:49.16--22:59:45.5 & 120 $\pm$ 11 &  27 & -0.32 $\pm$ 0.10  \\
284 & 042 & 18:02:49.24--22:50:18.7 &  71 $\pm$ 10 &  11 & -0.41 $\pm$ 0.11  \\
285 & 025 & 18:02:49.36--22:48:32.6 &  93 $\pm$ 12 &  10 &  0.11 $\pm$ 0.07  \\
286 & 041 & 18:02:49.66--22:48:55.6 & 100 $\pm$ 12 &  12 & -0.59 $\pm$ 0.07  \\
287 & 098 & 18:02:49.84--22:58:49.6 &  26 $\pm$  5 &   9 & -0.77 $\pm$ 0.25  \\
288 & 142 & 18:02:50.12--23:00:13.4 &  12 $\pm$  4 &   3 & -0.75 $\pm$ 0.31  \\
289 & 056 & 18:02:50.36--22:47:41.0 &  37 $\pm$ 10 &   4 &  0.20 $\pm$ 0.09  \\
290 & 034 & 18:02:50.52--22:48:50.3 &  93 $\pm$ 12 &  11 & -0.55 $\pm$ 0.07  \\
291 & 134 & 18:02:51.37--23:02:07.4 &  16 $\pm$  4 &   4 &  0.71 $\pm$ 0.23  \\
292 & 118 & 18:02:51.71--23:02:34.3 & 184 $\pm$ 15 &  26 & -0.60 $\pm$ 0.08  \\
293 & 050 & 18:02:52.27--22:49:48.8 &  18 $\pm$  6 &   3 &  0.03 $\pm$ 0.13  \\
294 & 150 & 18:02:53.09--23:03:39.0 &  22 $\pm$  6 &   5 &  0.38 $\pm$ 0.15  \\
295 & 149 & 18:02:54.90--22:58:22.1 &  14 $\pm$  4 &   4 &  0.83 $\pm$ 0.27  \\
296 & 023 & 18:02:57.08--22:54:20.1 & 168 $\pm$ 14 &  22 & -0.67 $\pm$ 0.09  \\
297 & 153 & 18:02:57.55--23:02:54.4 &  21 $\pm$  6 &   4 & -0.46 $\pm$ 0.18  \\
298 & 148 & 18:02:57.83--23:00:46.4 &  13 $\pm$  5 &   3 &  0.24 $\pm$ 0.22  \\
299 & 141 & 18:02:58.06--22:56:02.4 &  13 $\pm$  4 &   3 &  0.17 $\pm$ 0.21  \\
300 & 033 & 18:03:00.19--22:52:55.7 &  89 $\pm$ 12 &  11 & -0.47 $\pm$ 0.10  \\
301 & 147 & 18:03:01.61--23:00:53.6 &  25 $\pm$  6 &   5 & -0.29 $\pm$ 0.16  \\
302 & 152 & 18:03:02.19--23:02:48.7 &  45 $\pm$  9 &   6 &  0.58 $\pm$ 0.13  \\
303 & 146 & 18:03:03.32--23:00:19.2 &  25 $\pm$  7 &   4 & -0.19 $\pm$ 0.15  \\
304 & 133 & 18:03:03.93--23:01:04.2 &  79 $\pm$ 11 &  10 & -0.58 $\pm$ 0.11  \\
\enddata
\end{deluxetable}

\begin{deluxetable}{llllccccccccc}
\tabletypesize{\scriptsize}
\rotate
\tablenum{2}
\tablewidth{0pt}
\tablecaption{Summary of selected X-ray sources}\label{Tinter}
\tablehead{
\colhead{No\tablenotemark{a}} &
\colhead{WID\tablenotemark{a}} &
\colhead{CXOM20} &
\colhead{Net\tablenotemark{b}} &
\colhead{SNR\tablenotemark{b}} &
\colhead{max\tablenotemark{c}} &
\colhead{min\tablenotemark{c}} &
\colhead{ratio\tablenotemark{c}} &
\colhead{Prob.\tablenotemark{d}} &
\colhead{hardness\tablenotemark{e}} &
\colhead{2mass\tablenotemark{f}} &
\colhead{opti\tablenotemark{g}} &
\colhead{L$_x$} \\
\colhead{} &
\colhead{} &
\colhead{} &
\colhead{} &
\colhead{} &
\colhead{cts ks$^{-1}$} &
\colhead{cts ks$^{-1}$} &
\colhead{} &
\colhead{const} &
\colhead{ratio} &
\colhead{} &
\colhead{} &
\colhead{(10$^{31}$)} 
}
\startdata
   8& 196 &  18:01:59.07-22:52:27.4 &  73 &  12 &  5.87$\pm$2.26 &
1.44$\pm$0.34  &  4.07$\pm$1.84 &    55\%&   -0.49$\pm$0.11&     Y  &
[Y]  &0.32  \\
   23& 173 &  18:02:07.77-22:55:33.8 &  58 &  19 &  5.52$\pm$2.09 &
0.95$\pm$0.28  &  5.79$\pm$2.77 &     4\%&   -0.58$\pm$0.15&    TTS &
[Y]  &0.27  \\
38 (TC4)& 347 &  18:02:12.47-23:05:48.0 &  62 &   9 &  2.34$\pm$1.92 &
1.06$\pm$0.35  &  2.21$\pm$1.96 &   100\%&    0.67$\pm$0.13&     Y? &
N   &1.9  \\
91 (C2)&235  &18:02:23.05-23:01:58.1    &857  &129 &15.99$\pm$3.05  &
  9.72$\pm$0.52 &  1.65$\pm$0.33   &  2\% &  -0.47$\pm$0.03  &  N    &
  N & 6$^{h}$ \\
94 (CD) & 234 &  18:02:23.16-23:02:00.2 & 831 & 119 & 15.19$\pm$2.99 &
 6.79$\pm$2.22  &  2.24$\pm$0.85 &    79\%&   -0.50$\pm$0.03&     Y  &
 [Y] & 14$^{h}$ \\
    97& 272 &  18:02:23.25-23:01:34.9 &  48 &  12 &  3.15$\pm$1.76 &
1.03$\pm$0.29  &  3.06$\pm$1.91 &     8\%&   -0.18$\pm$0.13&    YSO &
N    &0.22 \\ 
102 (A) &230 &  18:02:23.54-23:01:50.9 & 884 & 114 & 19.95$\pm$3.34 &
 9.88$\pm$3.89  &  2.02$\pm$0.87 &    46\%&   -0.89$\pm$0.04&     Y  &
 [Y] &2 \\
106 (B)& 229 &  18:02:23.71-23:01:45.4 & 299 &  47 &  8.34$\pm$2.40 &
 3.15$\pm$1.76  &  2.65$\pm$1.67 &   100\%&   -0.64$\pm$0.07&     Y  &
 Y   & 5\\
 117 (TC1)& 264 &  18:02:25.16-23:01:26.7 &  41 &  15 &  2.37$\pm$1.63 &
0.69$\pm$0.27  &  3.44$\pm$2.72 &   100\%&    0.24$\pm$0.16&     N  &
N   & 4.7   \\
  166 & 212 &  18:02:30.37-22:59:29.0 & 135 &  57 &  8.78$\pm$2.43 &
2.26$\pm$0.32  &  3.88$\pm$1.21 &     0\%&   -0.19$\pm$0.09&    YSO &
U    &0.62 \\
  170 & 211 &  18:02:30.93-23:02:35.6 &  73 &  21 &  4.29$\pm$1.95 &
1.27$\pm$0.30  &  3.37$\pm$1.73 &    68\%&   -0.79$\pm$0.15&     N  &
Y    &0.33  \\
  194& 291 &  18:02:32.75-23:04:21.0 &  74 &  14 &  5.71$\pm$2.18 &
1.25$\pm$0.32  &  4.56$\pm$2.08 &     2\%&   -0.58$\pm$0.13&
(YSO)&    [Y]  &0.34   \\
  211&  72 &  18:02:34.51-22:59:37.9 & 312 & 107 &  9.98$\pm$2.55 &
2.77$\pm$1.69  &  3.60$\pm$2.38 &    43\%&   -0.47$\pm$0.06&      Y &
[Y]   &1.1  \\
  237&  29 &  18:02:38.85-22:50:06.0 &  81 &  13 &  5.57$\pm$2.20 &
1.79$\pm$0.35  &  3.12$\pm$1.38 &     2\%&    0.24$\pm$0.09&
(YSO)&     N   &6.32   \\
  246&  63 &  18:02:40.42-22:57:53.3 &  58 &  23 &  4.37$\pm$1.93 &
1.01$\pm$0.28  &  4.31$\pm$2.25 &    57\%&   -0.86$\pm$0.17&      Y &
[Y]   &0.25    \\
  256& 125 &  18:02:42.36-23:04:35.7 &  72 &  13 & 13.82$\pm$4.52 &
1.55$\pm$0.34  &  8.90$\pm$3.51 &     0\%&   -0.17$\pm$0.11&      N &
N     &0.33 \\
  283&  76 &  18:02:49.16-22:59:45.4 & 120 &  27 &  9.46$\pm$2.53 &
2.01$\pm$0.33  &  4.72$\pm$1.48 &     3\%&   -0.32$\pm$0.10&     YSO&
[Y]    &0.55  \\
  285 & 25  &  18:02:49.36-22:48:32.5 & 93 &  10 & 10.05$\pm$2.80 &
2.71$\pm$0.44 &  3.70$\pm$1.19  &  2\%  & 0.11$\pm$0.07 &    N   &
  N  & 6.2\\

\enddata
\tablenotetext{a}{X-ray source numbers are from Table 1, and
in the parenthesis, components A-D of HD164492 
and TC1 and TC4 (massive protostellar cores from \cite{cer98}) are marked.
WID is source number identified by wavelet analysis.}
\tablenotetext{b}{The net counts, and SNR is
detection signal to noise obtained from wavelet analysis}
\tablenotetext{c}{The count rate at the maxinum count in the bin and 
at the mininuum
count in the bin (see the text for detail),
and the ratio is the ratio between the two.}
\tablenotetext{d}{Probability of constancy which is estimated using
$\chi^2$  test. }
\tablenotetext{e}{Hardness ratio: spectral hardness ratio = (H-S)/(H+S) where
H and S is the count rate for hard (2-8 keV) and soft (0.5-2 keV) bands,
respectively.}
\tablenotetext{f}{TTS and YSO are identified using JHK color diagram
(details are described in Rho et al (2001)). Y: 2MASS detection, N:
2MASS non-detection. The identification with the parenthesis is uncertain
classification due to
uncertain 2MASS photometry.}
\tablenotetext{g}{Optical counterparts. Y: detection, the sources in
the Guide star catalog are marked []. N:non-detection. U: uncertain.}
\tablenotetext{h}{14$\times 10^{31}$ and 6$\times 10^{31}$ erg s$^{-1}$
are soft and hard components, respectively, for CD-C2 blend spectrum.}
\end{deluxetable}

\normalsize

\newpage
\clearpage
\begin{figure}
{Fig. 1 is a color figure which is available in jpeg format.}
\caption{Three color \chandra\ ACIS-I image of the Trifid Nebula.
The red, green and blue represents the images of 0.5-1 keV, 1-2 keV
and 2-8 keV, respectively. The image is centered at
R.A.\ $18^{\rm h} 02^{\rm m} 26^{\rm s}$ and Dec.\
$-23^\circ$00$^{\prime} 39^{\prime \prime}$ (J2000) and covers 7.5
arcmin field of view. }
\label{chandrazoom}
\end{figure}

\begin{figure}
\psfig{figure=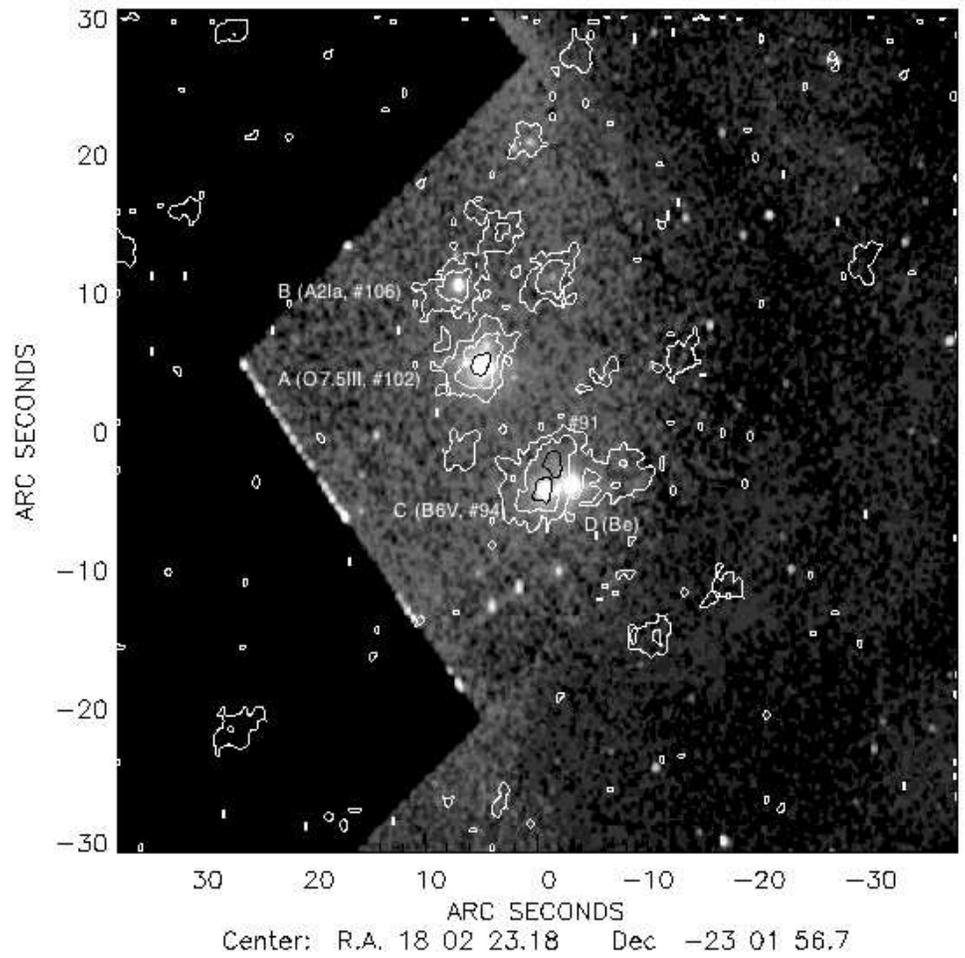,height=17.5truecm,angle=0}
\caption{Sources in HD164492 complex.
The gray scale image of HST superposed on \chandra\ X-ray contours. The
components of HD164492 are marked. }
\label{ostarcomplex}
\end{figure}

\begin{figure}
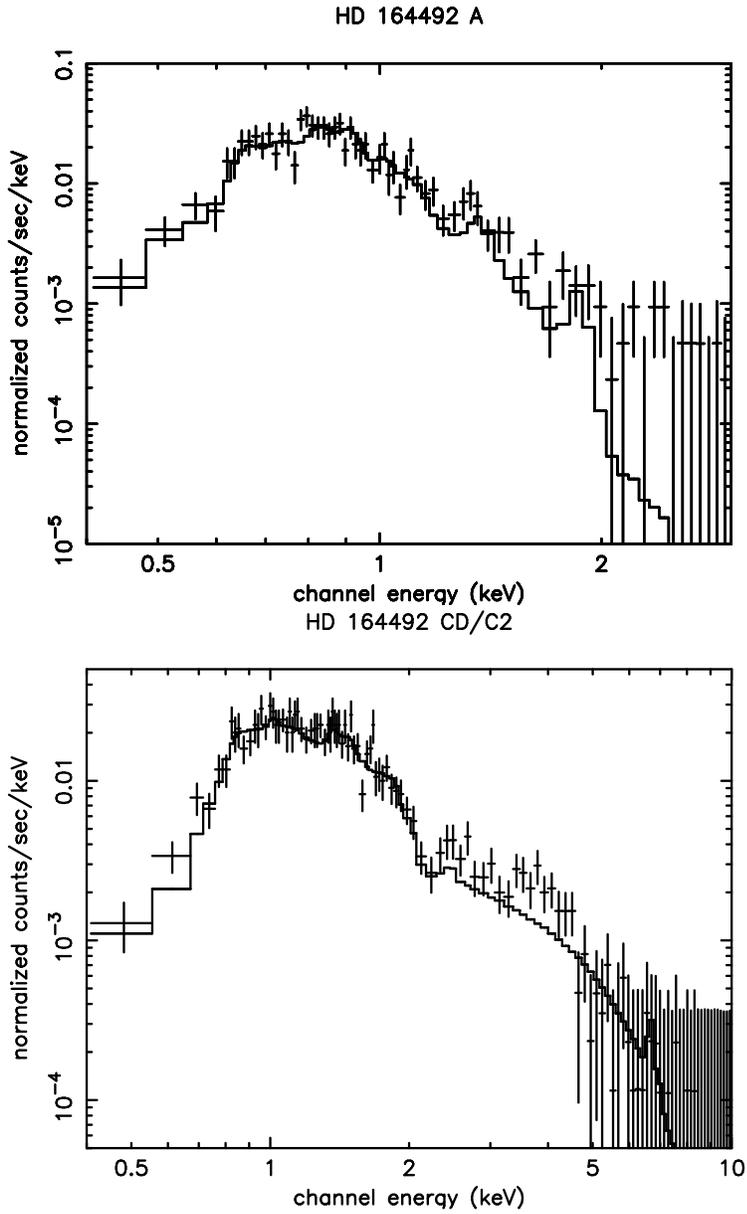

\vbox{{
\psfig{figure=f3a.ps,height=8truecm,angle=270}
\psfig{figure=f3b.ps,height=8truecm,angle=270}
}}
\caption{X-ray spectra of HD 164492 A, O7.5III star (a) and HD 164492 CD-blend,
B/Be stars (b) }
\label{ostarspec}
\end{figure}

\begin{figure}[ht]
\psfig{figure=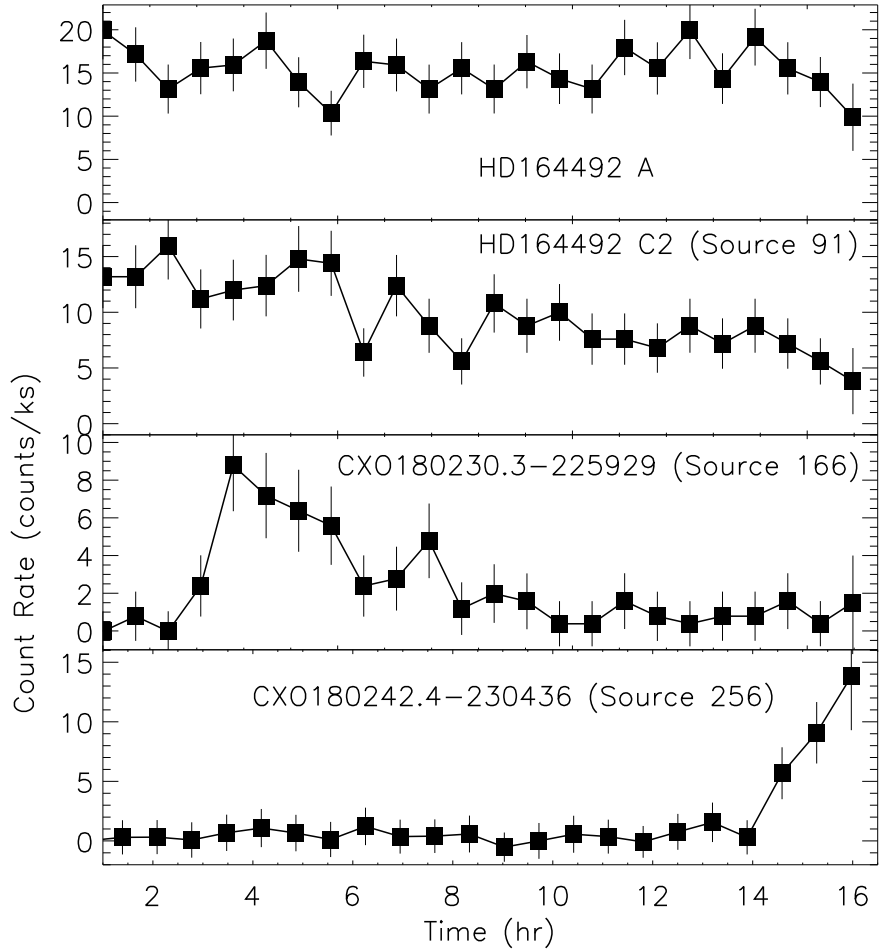,height=13truecm,angle=0}
\caption{Light curves from HD164492 A and C2, and two pre-main-sequence stars.
The source numbers are marked for each (see Table 1)}
\label{lcpms}
\end{figure}

\begin{figure}[ht]
\psfig{figure=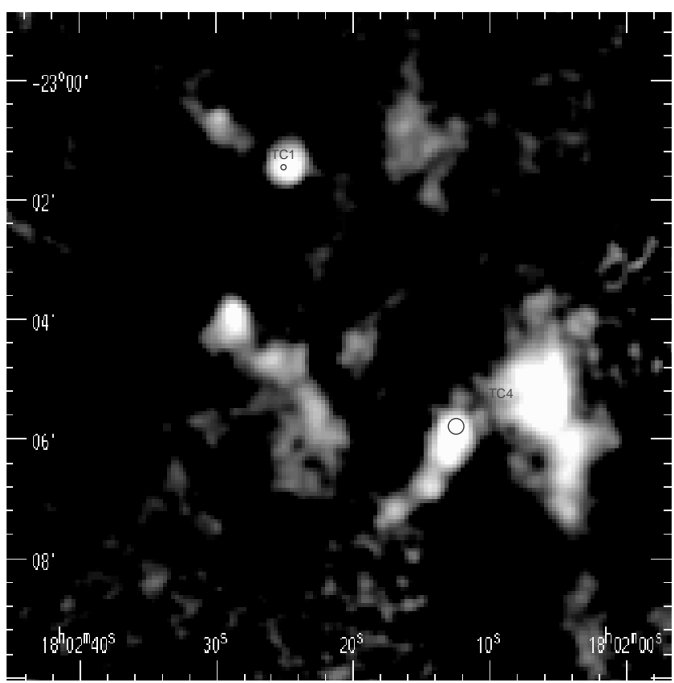,height=10truecm,angle=0}
\caption{Thirteen hundred-micrometer dust continuum map \citep{cer98}
and two X-ray sources coinciding with two of the earliest protostellar cores
of TC1 and TC4.}
\label{tc1tc4}
\end{figure}

\end{document}